\documentclass[aps,showpacs,twocolumn,longbibliography]{revtex4-1}
\usepackage[utf8]{inputenc}
\usepackage{amsfonts}
\usepackage{amssymb}
\usepackage{amsmath}
\usepackage{graphicx}
\usepackage{epsfig}
\usepackage{subfigure}
\usepackage{appendix}
\usepackage{color}
\usepackage{hyperref}
\usepackage{fullpage}

\hypersetup{hypertex=true,
colorlinks=true,
linkcolor=blue,
anchorcolor=blue,
urlcolor=blue,
citecolor=blue}
\begin{document}

\title{Periodic dynamics in an Ising chain with a quadratic transverse field}
\author{H. P. Zhang}
\author{Z. Song}
\email{songtc@nankai.edu.cn}

\begin{abstract}
A quadratic well plays a central role in a wide variety of modern physical
theories and applications. In this work, we investigate many-body dynamics
in a quadratic well, using an Ising chain as a paradigmatic example. In
contrast to a uniform Ising chain, where the quantum phase transition is
driven by the field strength, the present system exhibits spatially varying
quantum phases along the chain. Through analysis of the Majorana
representation, we obtain exact solutions for localized modes, revealing a
topologically degenerate spectrum in the thermodynamic limit. In the case of
a finite-size quantum phase region, the Kramers-like degeneracy is lifted by
a constant shift, leading to periodic oscillations for a finite-temperature
thermal initial state. Numerical simulations of the magnetization, local
density of state, and quench fidelity support our conclusions. Our findings
enrich the understanding of many-body dynamics in a trapping field.
\end{abstract}

\affiliation{School of Physics, Nankai University, Tianjin 300071, China}

\maketitle
\section{Introduction}

\label{Introduction}

The quadratic well is not just an idealized concept; it is an essential tool
in condensed matter physics for modeling real-world phenomena. It simplifies
complex systems into tractable models, allowing for the study of a wide
range of phenomena, including thermal dynamics, electron behavior, and
quantum effects in condensed matter. For instance, lattice vibrations are
modeled as harmonic oscillators, enabling the concept of phonon
quasiparticles that carry heat and sound. Quantum dots and semiconductor
heterostructures often use quadratic confinement potentials. Ultracold atoms
in laser traps experience approximately quadratic potentials, crucial for
quantum simulation. The corresponding theoretical approaches mainly focused
on the model describing non-interacting particles in a trapping field.
Recent advancements in the field of condensed matter physics and quantum
simulation have enabled researchers to explore novel quantum phenomena in
controlled environments. For instance, ultracold atoms in optical lattices
have emerged as a powerful platform for simulating complex quantum systems,
allowing for the manipulation of interactions and external fields with
unprecedented precision . This has opened up new possibilities for
investigating the dynamics of many-body system in trapping well. Novel
phenomena may emerge from the interplay between internal field profiles and
particle interactions.

In this work, In this work, we delve into the dynamics of an Ising chain in
a quadratic transverse field well. The Ising model with a uniform transverse
field is a paradigmatic example of quantum phase transitions and is often
studied in the context of quantum computing \cite%
{Pfeuty1970,Sachdev2011,Dutta2015}. At the quantum critical point, the
system undergoes a quantum phase transition, which is characterized by a
sudden change in the ground state and in the behavior of correlation
functions. Experimentally, the transverse field Ising chain has direct
realizations \cite{Bitko1996, Coldea2010}, and is also achievable with
ultracold atoms in optical lattices \cite{Simon2011, Islam2011}. The
identification of quantum phase diagrams holds vital importance for both
condensed matter physics and quantum information science. Over the past few
decades, this field has witnessed extensive theoretical and experimental
investigations~\cite{Fisher1990, Bitko1996, Vojta2000, Si2001, Porras2004,
Uhlarz2004, Roennow2005, Coldea2010, Kim2011, Simon2011, Trenkwalder2016,
Rem2019}. Our investigation is motivated by the intriguing question of how
the dynamics is manifested when the transverse field is non-uniform.
Although this Hamiltonian is not exactly solvable, the Jordan-Wigner and
Majorana transforms are still applicable.\ Through analysis of the Majorana
representation, we obtain exact solutions for localized modes, revealing a
topologically degenerate spectrum in the thermodynamic limit. There are two
localized modes, which reside at the interfaces of three regions associated
with three quantum phases. We find that when the middle region is finite in
size, the Kramers-like degeneracy is lifted by a constant shift due to the
hybridization of the two localized modes. This Zeeman effect leads to
periodic oscillations for a finite-temperature thermal initial state.
Numerical simulations of the magnetization and local density of state
distributions along the chain map out the phase diagram in real space.
Additionally, we compute quench dynamics to corroborate our conclusions. Our
findings not only address the many-body system itself, but also predict
finite-temperature phenomena, thereby enriching our understanding of
many-body dynamics in a trapping field.

This paper is organized as follows. In Sec. \ref{Model and solution}, we
introduce the quadratic transverse-field Ising chain and map it to a
Majorana lattice via the Jordan-Wigner transformation. Sec. \ref{Localized
modes and topological degeneracy} is dedicated to analyzing the Gaussian
zero modes localized at the interfaces of quantum phases and their role in
inducing a Kramers-like topological degeneracy. In Sec. \ref{Scaling
behavior}, we investigate the quantum phase transitions using three
quantities: local density of states, magnetization, and susceptibility. In
Sec. \ref{Thermal state oscillation}, we explore the dynamical signature of
the quasi-degenerate spectrum via quench dynamics from a thermal state.
Finally, we provide a summary in Sec. \ref{Summary}.

\section{Model and solution}

\label{Model and solution}

\begin{figure}[t]
\centering
\includegraphics[width=0.95\linewidth]{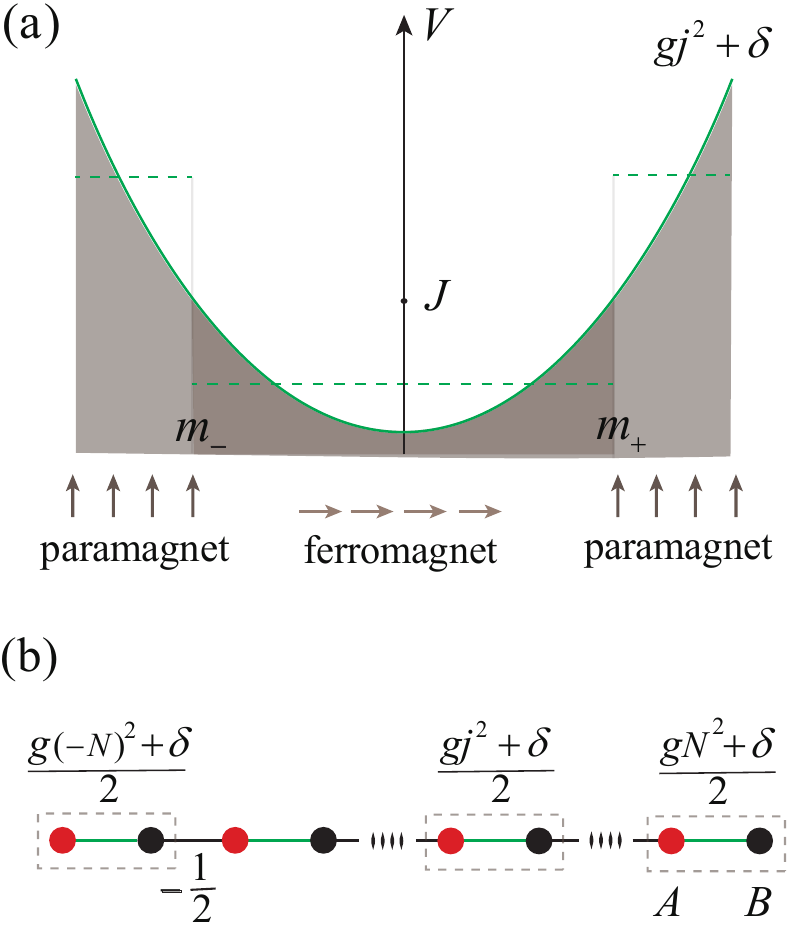}
\caption{(a) Schematic illustrations of the Hamiltonian in Eq.~(\protect\ref%
{H}), which represents a Ising chain with quadratic transverse field (solid
line). Here, $-J$ is the coupling strength between adjacent spins, $g$ and $%
\protect\delta $ denote the strength and baseline of the parabola,
respectively. The phase boundaries $m_{\pm }$ separate the system into a
ferromagnetic region $(gj^{2}+\protect\delta <J)$ and paramagnetic regions $%
(gj^{2}+ \protect\delta >J)$. The nature of the ground state can be
approximately the same as that of a system consisting of three uniform
segments (dashed line). (b) The Majorana lattice defined in Eq.~(\protect\ref%
{h_M}), with intracell and intercell hopping strength of half the quadratic
field and $-1/2$, respectively. It features two robust zero modes localized
around the unit cells $m_{\pm }$ in the single-particle invariant subspace,
as illustrated in Fig.~\protect\ref{fig:fig2}. We adopt $J=1$\ as the
unified energy scale.}
\label{fig:fig1}
\end{figure}

We consider an Ising chain of length $L=2N+1$ with a transverse field being
a quadratic function of position. The Hamiltonian is given by%
\begin{equation}
H=-J\sum_{j=-N}^{N-1}\sigma _{j}^{x}\sigma _{j+1}^{x}+\sum_{j=-N}^{N}\left(
gj^{2}+\delta \right) \sigma _{j}^{z},  \label{H}
\end{equation}%
where $\sigma _{j}^{\alpha }$\ ($\alpha =x,y,z$) are the Pauli operators on
site $j$, $g$\ ($g\geq \left( J-\delta \right) /N^{2}$) is the quadratic
transverse-field strength, and $\delta $ ($0<\delta <J$) is the baseline of
the uniform transverse field. For simplicity, we assume $J=1$ and let the
site index $j$ range from $-N$ to $N$. The system is schematically
illustrated in Fig. \ref{fig:fig1}(a). It can be checked that the model
respects the parity symmetry, that is, the parity operator $%
p=\prod_{j=-N}^{N}\left( -\sigma _{j}^{z}\right) $ commutes with the
Hamiltonian.

We start with a brief review of the uniform transverse-field Ising model ($%
g=0$) under both periodic and open boundary conditions, which has been
thoroughly investigated \cite{Elliott1970,Pfeuty1970,Zhang2021}. As a
paradigm of second-order QPT, the former is exactly solvable \cite%
{Pfeuty1970}. At zero temperature, the QPT occurs at $\delta =1$, separating
the ferromagnetic $\left( \delta <1\right) $ and paramagnetic $\left( \delta
>1\right) $\ phases. The latter, in the thermodynamic limit, exhibits an
additional symmetry within the ferromagnetic phase $\left( \delta <1\right) $%
, which is associated with the topological Kramers-like degeneracy \cite%
{Zhang2021}. In the case of nonzero $g$, the strength of the transverse
field varies spatially along the chain. We note that the resulting field
passes through the critical points twice. Intuitively, the nature of the
ground state can be approximately the same as that of a system consisting of
three uniform segments, separated by the sites $m_{\pm }=\pm \lfloor \sqrt{%
\left( 1-\delta \right) /g}\rfloor $, where $\lfloor \cdot \rfloor $\
denotes the floor function. The transverse field is larger than $1$\ in the
region $j\in \lbrack -N,m_{-})\cup (m_{+},N] $, while smaller than $1$\ in
the region $j\in \lbrack m_{-},m_{+}]$. Therefore, the three regions are in
the paramagnetic, ferromagnetic, and paramagnetic phases, respectively. This
is schematically illustrated in Fig.~\ref{fig:fig1}(a).

Although the exact solution cannot be obtained for the non-uniform chain
with $g\neq 0$, there is no doubt that, far from the positions $m_{\pm }$,
the system with large $N$\ resides in the paramagnetic, ferromagnetic, and
paramagnetic phases, respectively. The main question is what happens in the
vicinity of $m_{\pm }$. Some key characteristics of the system can be
revealed through the following procedure.

Taking the Jordan--Wigner transformation \cite{Jordan1928} 
\begin{eqnarray}
\sigma _{j}^{x} &=&\prod_{l<j}\left( 1-2c_{l}^{\dag }c_{l}\right) \left(
c_{j}+c_{j}^{\dag }\right) ,  \notag \\
\sigma _{j}^{z} &=&2c_{j}^{\dag }c_{j}-1,
\end{eqnarray}%
to replace the Pauli operators by the fermionic operators $c_{j}$, the
Hamiltonian~(\ref{H}) with $J=1$ is expressed as%
\begin{eqnarray}
H &=&-\sum_{j=-N}^{N-1}\left( c_{j}^{\dag }c_{j+1}+c_{j}^{\dag
}c_{j+1}^{\dag }\right) +\mathrm{H.c.}  \notag \\
&&+\sum_{j=-N}^{N}\left( gj^{2}+\delta \right) \left( 2c_{j}^{\dag
}c_{j}-1\right) .
\end{eqnarray}%
which is a Kitaev model \cite{Kitaev2001} with quadratic on-site potential.
To get the diagonal form of the Hamiltonian, we introduce the Majorana
fermion operators 
\begin{equation}
a_{j}=c_{j}^{\dag }+c_{j},b_{j}=-i\left( c_{j}^{\dag }-c_{j}\right) ,
\end{equation}%
which satisfy the commutation relations 
\begin{equation}
\left\{ a_{j},a_{j^{\prime }}\right\} =2\delta _{jj^{\prime }},\left\{
b_{j},b_{j^{\prime }}\right\} =2\delta _{jj^{\prime }},\left\{
a_{j},b_{j^{\prime }}\right\} =0.
\end{equation}%
Then the Majorana representation of the original Hamiltonian is%
\begin{eqnarray}
H &=&-\frac{i}{2}\sum_{j=-N}^{N-1}b_{j}a_{j+1}-\frac{i}{2}%
\sum_{j=-N}^{N}\left( gj^{2}+\delta \right) a_{j}b_{j}+\mathrm{H.c.}  \notag
\\
&=&\varphi ^{\dag }h_{\mathrm{M}}\varphi ,
\end{eqnarray}%
where $\varphi ^{\dag }=\left( a_{-N},ib_{-N},a_{-N+1},ib_{-N+1},\text{ }%
\cdots ,a_{N},ib_{N}\right) $. The core matrix $h_{\mathrm{M}}$ can be
explicitly written as%
\begin{eqnarray}
h_{\mathrm{M}} &=&\frac{1}{2}\sum_{j=-N}^{N}\,\left( gj^{2}+\delta \right)
|j\rangle _{A}\langle j|_{B}\,  \notag \\
&&-\frac{1}{2}\sum_{j=-N}^{N-1}\,|j\rangle _{B}\langle j+1|_{A}\,+\mathrm{%
H.c.,}  \label{h_M}
\end{eqnarray}%
where basis $\left\{ |l\rangle _{A},|l\rangle _{B}\right\} $ is an
orthonormal complete set, satisfying $\langle l|_{A}|l^{\prime }\rangle _{B}$
$=\delta _{ll^{\prime }}\delta _{AB}$. The physics of $h_{\mathrm{M}}$ is
very clear. It represents a $2L$-site generalized Su-Schrieffer-Heeger (SSH)
chain, referred to as the Majorana lattice, in the single-particle invariant
subspace. A schematic illustration of its structure is shown in Fig. \ref%
{fig:fig1}(b). We note that the tight-binding model corresponds to $h_{%
\mathrm{M}}$\ has chiral ($\mathcal{C}$) symmetry \cite%
{Guo2015,Kraus2016,Lin2017,Lee2016,Li2015,Malzard2015,Asboth2016}, 
\begin{equation}
\mathcal{C}h_{\mathrm{M}}\mathcal{C}^{-1}=-h_{\mathrm{M}},
\end{equation}%
where the operator $\mathcal{C}$\ is defined as 
\begin{equation}
\mathcal{C}\left\vert j\right\rangle _{\mathrm{A}}=\left\vert j\right\rangle
_{\mathrm{A}},\text{ }\mathcal{C}\left\vert j\right\rangle _{\mathrm{B}%
}=-\left\vert j\right\rangle _{\mathrm{B}}.
\end{equation}

The anticommutation relation between operators $\mathcal{C}$ and $h_{\mathrm{%
M}}$ leads to the equations%
\begin{eqnarray}
h_{\mathrm{M}}|\phi _{n}\rangle &=&\lambda _{n}|\phi _{n}\rangle ,  \notag \\
h_{\mathrm{M}}\mathcal{C}|\phi _{n}\rangle &=&-\lambda _{n}\mathcal{C}|\phi
_{n}\rangle ,
\end{eqnarray}%
which indicates that if $|\phi _{n}\rangle $ is an eigenvector of $h_{%
\mathrm{M}}$ with eigenvalue $\lambda _{n}$, then $\mathcal{C}|\phi
_{n}\rangle $ is also an eigenvector with eigenvalue $-\lambda _{n}$. This
allows us to rewrite the Majorana representation in terms of a free-fermion
representation 
\begin{equation}
H=\sum_{n=0}^{2N}\varepsilon _{n}\left( d_{n}^{\dag }d_{n}-\frac{1}{2}%
\right) ,
\end{equation}%
where $\varepsilon _{n}=4\left\vert \lambda _{n}\right\vert $ and $%
d_{n}=\varphi ^{\top }\phi _{n}/\sqrt{2}$ is a fermion operator satisfying $%
\left\{ d_{n},d_{n^{\prime }}\right\} =0$ and $\left\{ d_{n},d_{n^{\prime
}}^{\dag }\right\} =\delta _{nn^{\prime }}$. Here, $\phi _{n}$ is the
representation of $|\phi _{n}\rangle $ in the orthogonal basis $\left\{
|l\rangle _{A},|l\rangle _{B}\right\} $. The spectrum $\varepsilon _{n}$ and
the expression for the fermion operator$\ d_{n}$ can be obtained by exact
diagonalization of the matrix $h_{\mathrm{M}}$. The ground state $\left\vert
G\right\rangle $\ of $H$\ is the vacuum state of all the operator $d_{n}$,
i.e., $d_{n}\left\vert G\right\rangle =0$, with the ground state energy 
\begin{equation}
E_{g}=-\frac{1}{2}\sum_{n=0}^{2N}\varepsilon _{n},
\end{equation}%
where the set of numbers $\left\{ \varepsilon _{n}\right\} $\ are labelled
in ascending order such that $\varepsilon _{0}<\varepsilon
_{1}<...<\varepsilon _{2N}$. Accordingly, starting from the ground state $%
\left\vert G\right\rangle $, any eigenstate can be constructed by $%
\prod_{\left\{ n_{l}\right\} }d_{n_{l}}^{\dag }\left\vert G\right\rangle $,\
where $\left\{ n_{l}\right\} $ denotes a set of occupied single-particle
states, and the corresponding eigenenergy is $E_g+\sum_{\left\{ n_{l}\right\}
}\varepsilon _{n_{l}}$. 
\begin{figure*}[tbp]
\centering
\includegraphics[width=0.85\linewidth]{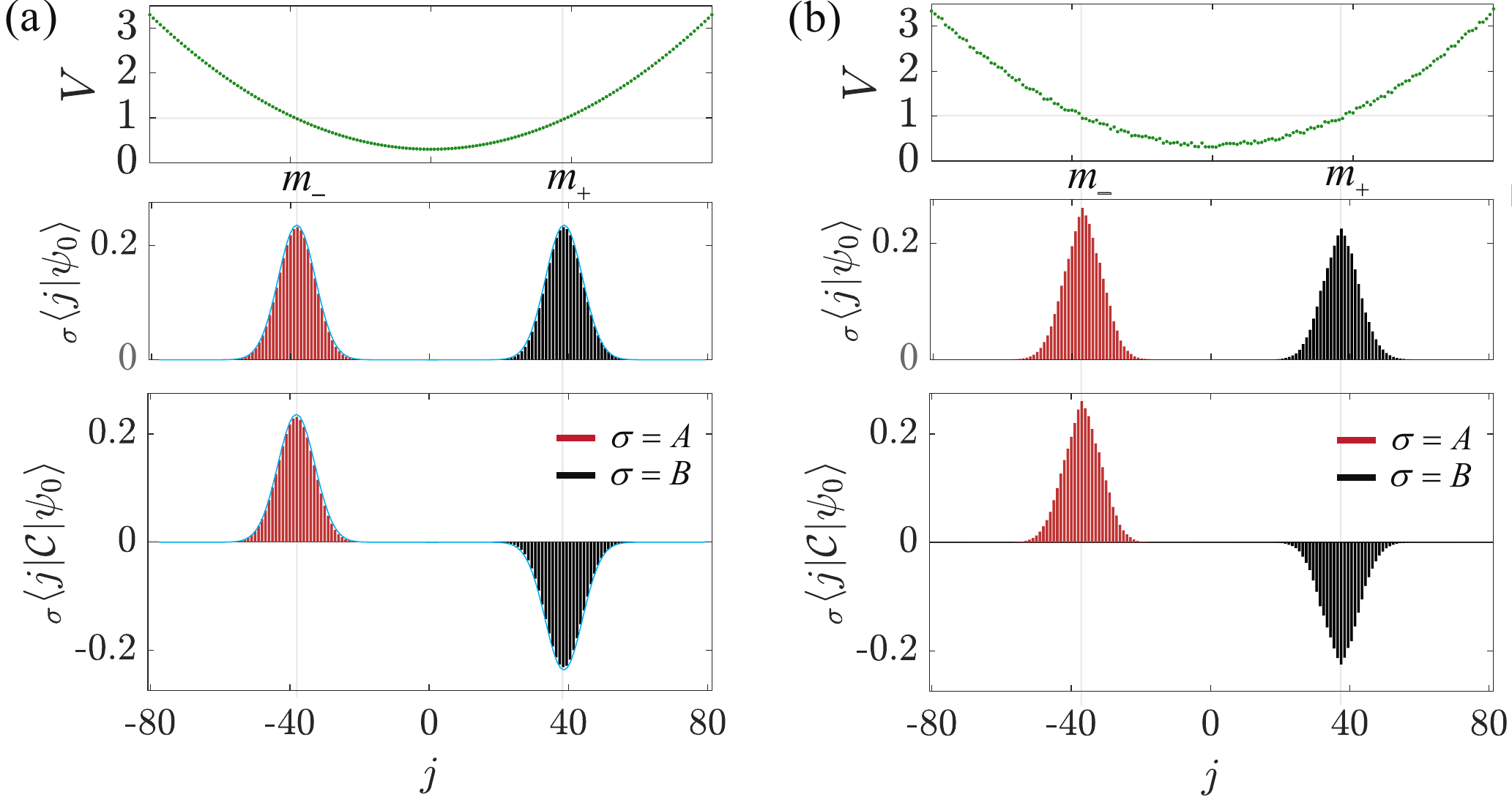}
\caption{ Numerical results for zero modes in the Majorana lattice $h_%
\mathrm{M}$ (defined in Eq.~(\protect\ref{h_M})) of Ising models under exact
and random disorder quadratic transverse fields. (a) In the exact case, the
results are consistent with our theoretical results in Eq.~(\protect\ref%
{Gauss}) (blue line), where the zero modes exhibit a Gaussian profile
centered at $m_\pm$. (b) In the case of random disorder, robust zero modes
centered at $m_\pm$ are still observed. According to Eq.~(\protect\ref{DPsi}%
), the zero modes imply a topological degeneracy in Hamiltonian (\protect\ref%
{H}) that is immune to perturbations. Other parameters: $N=80$, $g
=5\times10^{-4}$, $\protect\delta =0.3$, and $J=1$.}
\label{fig:fig2}
\end{figure*}

\section{Localized modes and topological degeneracy}

\label{Localized modes and topological degeneracy}

It has been well established that the zero modes of the Majorana lattice for
a uniform chain play a crucial role in characterising the topological
features of the quantum phase \cite{Zhang2021,Kitaev2001,Nayak2008}. From
the point of view of quantum dielectric theory, the zero modes can be
regarded as localized charges residing at the interface between two phases 
\cite{Levin2012,Read2012,Lutchyn2010,Alicea2012}. It is presumable that
there exist localized modes in the vicinity of the positions $m_{\pm }$ for
the present Hamiltonian $H$.

In this section, we investigate the localized modes of the Majorana lattice
described by $h_{\mathrm{M}}$ and their effects on the properties of $H$. We
start with a pair of Bethe ansatz eigenvectors of the form 
\begin{equation}
|\phi _{A}\rangle =\sum_{j=-N}^{m_{+}}\alpha _{j}|j\rangle
_{A},\,\,\,\,|\phi _{B}\rangle =\sum_{j=m_{-}}^{N}\alpha _{-j}|j\rangle _{B},
\end{equation}%
where the coefficients are given by 
\begin{equation}
\alpha _{j}=\frac{1}{\sqrt{\Omega }}\prod\limits_{k=-N-1}^{j-1}\left(
gk^{2}+\delta \right) ,  \label{apha_j}
\end{equation}%
with the normalization factor $\Omega
=\sum_{j=-N}^{m_{+}}\prod_{k=-N-1}^{j-1}\left( gk^{2}+\delta \right) ^{2}$.
Applying the matrix $h_{\mathrm{M}}$\ to the vectors $|\phi _{A}\rangle $\
and $|\phi _{A}\rangle $, we have%
\begin{equation}
\left\vert h_{\mathrm{M}}|\phi _{A}\rangle \right\vert =\left\vert h_{%
\mathrm{M}}|\phi _{B}\rangle \right\vert =\frac{\Delta }{2}\alpha _{m_{-}}<%
\frac{\Delta }{2},
\end{equation}%
where $\Delta =\prod_{k=m_{-}}^{m_{+}}\left( gk^{2}+\delta \right) $ is a
function of $g$ and $\delta $, but is independent of $N$. When $g\ll 1$, $%
\Delta $ is very small and tends to zero as $m_{-}${} approaches infinity.
Obviously, this requires the chain to be in the thermodynamic limit, $%
N\rightarrow \infty $.

From Eq.~(\ref{apha_j}), we note that $\alpha _{j}$\ reaches its maximum at $%
j=m_{-}$.\ In the vicinity of $m_{-}$, we have%
\begin{eqnarray}
\frac{\text{\textrm{d}}}{\text{\textrm{d}}j}\ln \alpha _{j} &\approx &\ln
\alpha _{j+1}-\ln \alpha _{j}=\ln \left( gj^{2}+\delta \right)  \notag \\
&\approx &-2\sqrt{(1-\delta )g}(j-m_{-}),
\end{eqnarray}%
by using the Taylor expansion for $g\ll 1$. We then have the approximate
expression of the localized mode%
\begin{equation}
\alpha _{\pm j}\propto \exp [-\sqrt{(1-\delta )g}(j-m_{\mp })^{2}],
\label{Gauss}
\end{equation}%
which is of Gaussian type. They are also referred to as zero modes since we
have vanishing $\Delta $\ for sufficiently large $m_{\pm }$.

In contrast, in the case with moderate $m_{\pm }$, a finite $\Delta $
results in the hybridization between $|\phi _{A}\rangle $ and $|\phi
_{B}\rangle $, which in turn induces a splitting of the zero-eigenvalue
degeneracy. Within the subspace spanned by $\left\{ |\phi _{A}\rangle ,|\phi
_{B}\rangle \right\} $, the matrix elements of $h_{\mathrm{M}}$ are given by 
\begin{eqnarray}
\langle \phi _{A}|h_{\mathrm{M}}|\phi _{B}\rangle &=&\langle \phi _{B}|h_{%
\mathrm{M}}|\phi _{A}\rangle =\lambda _{0},  \notag \\
\langle \phi _{A}|h_{\mathrm{M}}|\phi _{A}\rangle &=&\langle \phi _{B}|h_{%
\mathrm{M}}|\phi _{B}\rangle =0,
\end{eqnarray}%
with $\lambda _{0}=\Delta \alpha _{m_{-}}^{2}/2$. Therefore, the hybridized
eigenvectors are $|\phi _{0}\rangle =\left( |\phi _{A}\rangle +|\phi
_{B}\rangle \right) /\sqrt{2}$, and $\mathcal{C}|\phi _{0}\rangle =\left(
|\phi _{A}\rangle -|\phi _{B}\rangle \right) /\sqrt{2}$, respectively, with
the corresponding eigenvalue $\pm \lambda _{0}$.

Now we turn to the effects of the localized modes on the ground-state
properties of $H$. Based on the expression of $|\phi _{0}\rangle $, the
corresponding fermion operator can be expressed as%
\begin{equation}
d_{0}=\frac{1}{2}\sum_{j=-N}^{m_{+}}\alpha _{j}\left( c_{j}^{\dag
}+c_{j}-c_{-j}^{\dag }+c_{-j}\right) ,
\end{equation}%
with a small value of the energy $\varepsilon _{0}=4\lambda _{0}=2\Delta
\alpha _{m_{-}}^{2}$. The localized form of the profile of $\alpha _{j}$\
results in the localization of the operator $d_{0}$ around the position $%
m_{\pm }$. Furthermore, applying the inverse Jordan-Wigner transformation, $%
d_{0}$ can be expressed as the combination of spin operators,%
\begin{equation}
d_{0}=\frac{1}{2}\sum_{j=-N}^{m_{+}}\alpha _{j}\tau _{j},
\end{equation}%
where $\tau _{j}=\prod_{l<j}\left( -\sigma _{l}^{z}\right) \sigma
_{j}^{x}-i\prod_{l<-j}\left( -\sigma _{l}^{z}\right) \sigma _{-j}^{y}$ is
position-dependent operator. Although the exact eigenstates cannot be
obtained, the two operators $p$ and $d_{0}$\ can be used to reveal their
structure. If $\left\{ |\psi _{n}^{+}\rangle \right\} $ is a set of
eigenstates of $H$ with eigenenergies $\left\{ E_{n}^{+}\right\} $, i.e., $%
H|\psi _{n}^{+}\rangle =E_{n}^{+}|\psi _{n}^{+}\rangle $ and $p|\psi
_{n}^{+}\rangle =|\psi _{n}^{+}\rangle $, there exists another set of
eigenstates $\left\{ |\psi _{n}^{-}\rangle \right\} $\ of $H$ with
eigenenergies $\left\{ E_{n}^{-}\right\} $, which satisfy $p|\psi
_{n}^{-}\rangle =-|\psi _{n}^{-}\rangle $, and 
\begin{equation}
|\psi _{n}^{-}\rangle =d_{0}|\psi _{n}^{+}\rangle
,\,\,E_{n}^{-}=E_{n}^{+}-\varepsilon _{0}.  \label{DPsi}
\end{equation}
When $\varepsilon _{0}\rightarrow 0$, a degeneracy emerges in the system.
This is referred to as topological Kramers-like degeneracy \cite{Zhang2021},
as it persists under random, position-dependent perturbations of the field $%
gj^{2}+\delta $. This robustness originates from the localization of the
zero modes associated with the corresponding $h_{\mathrm{M}}$. The profiles
of the localized modes $|\psi _{0}\rangle $ and $\mathcal{C}|\psi
_{0}\rangle $, obtained by exact diagonalization of a finite system, are
plotted in Fig. \ref{fig:fig2}. These profiles are in accordance with our
prediction.

\begin{figure*}[tbp]
\centering
\includegraphics[width=0.95\linewidth]{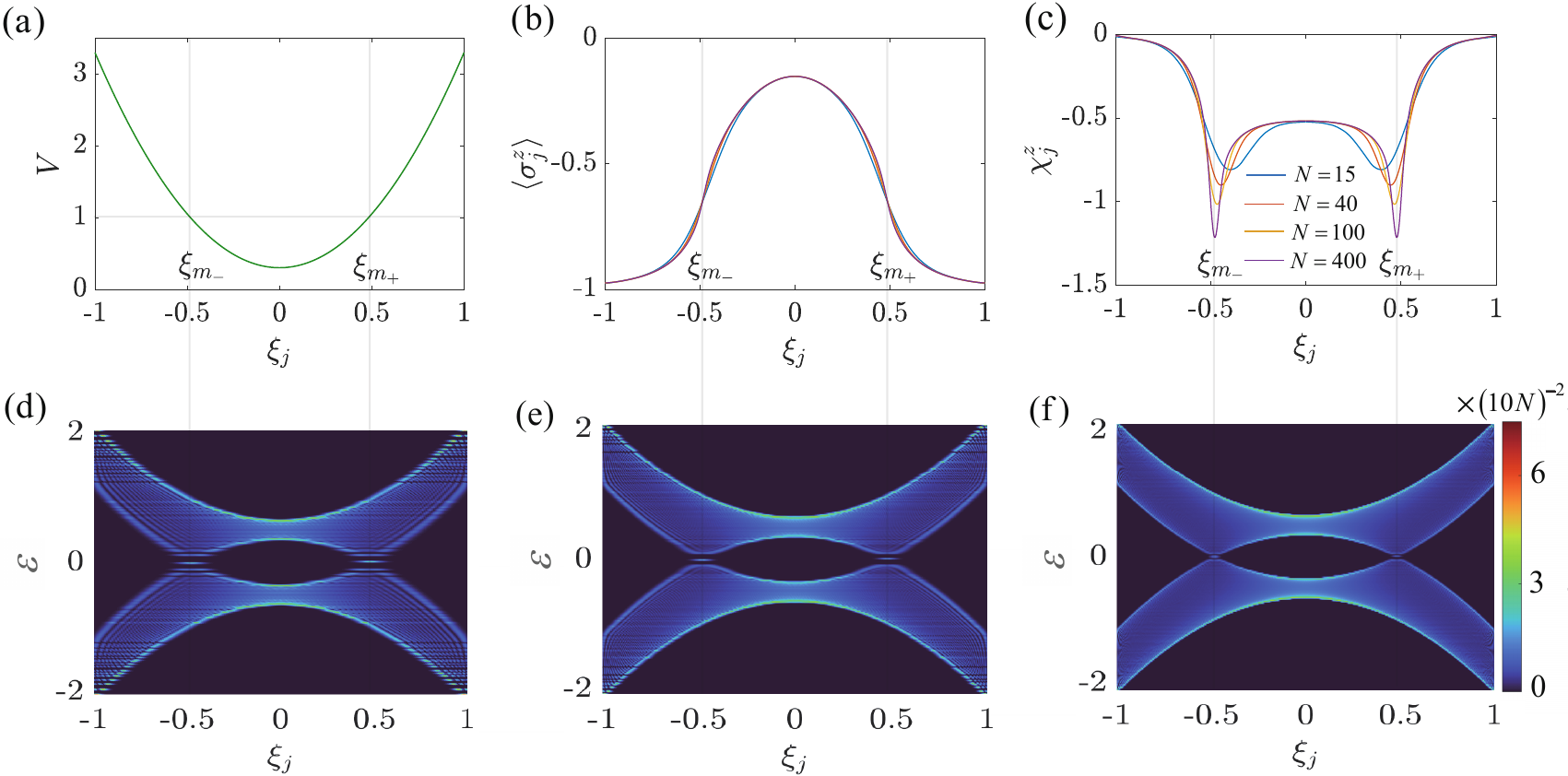}
\caption{Plots of the local density of states $D_{j}(\protect\varepsilon )$,
magnetizations $\left\langle \protect\sigma _{j}^{z}\right\rangle $ and
magnetic susceptibility $\protect\chi _{j}^{z}$ of the ground state, as
defined in Eqs.~(\protect\ref{D_j}), (\protect\ref{sigmaz}), and (\protect
\ref{chiz}), respectively. (a) Quadratic transverse field in the Hamiltonian
(\protect\ref{H_scale}) with parameters $g_{N}=3$, $\protect\delta =0.3$,
and $J=1$, which be used in (b-f). The $N$-dependent positions $\protect\xi %
_{m_{\pm }}$ separate the ferromagnetic and paramagnetic regions. (b) $%
\left\langle \protect\sigma _{j}^{z}\right\rangle $ for different $N$
intersect at the same points $\protect\xi _{m_{\pm }}$, and the crossing
becomes steeper as $N$ increases. (c) The peak of $\protect\chi _{j}^{z}$
becomes significantly sharper as $N$ increases and shifts toward $\protect%
\xi _{m_{\pm }}$, suggesting its divergence at $\protect\xi _{m_{\pm }}$ in
the thermodynamic limit and signaling a phase transition. (d-f) The local
density of state $D_{j}$\ for different $N$. The distributions of $D_{j}$
become more continuous as $N$ increases. The band gap reaches its minima at
points $\protect\xi _{m_{\pm }}$. Peaks appear in the mid-gap, indicating the localized
modes.}
\label{fig:fig3}
\end{figure*}

\section{Scaling behavior}

\label{Scaling behavior}

In general, QPTs are often investigated in uniform systems, while our study
focuses on examinyng this topic in systems with position-dependent
parameters. In addition, standard QPTs occur in the thermodynamic limit,
whereas finite-size systems exhibit scaling behavior. In this section, we
establish the connection between these two types of QPTs. To this end, we
rewrite the Hamiltonian in Eq. (\ref{H}) in the form

\begin{equation}
H=-J\sum_{j=-N}^{N-1}\sigma _{j}^{x}\sigma _{j+1}^{x}+\sum_{j=-N}^{N}\left(
g_{N}\xi _{j}^{2}+\delta \right) \sigma _{j}^{z},  \label{H_scale}
\end{equation}%
where the $N$-dependent position and field strength are given by $\xi
_{j}=j/N$ and $g_{N}=N^{2}g$, respectively. This allows us to perform the
following analysis.

We start with the case in which $g_{N}$ is constant for any given $N$. For
sufficiently large $N$, within the region of the chain, $\xi _{j}$ is
approximately constant and the slope of the transverse field approximately
vanishes. Within this region, the subsystem can be regarded as a uniform
chain, to which the standard QPT theory can be applied. Accordingly, a small
shift of $\xi _{j}$ corresponds to a slight change of the uniform external
field. In this sense, all the features of a uniform Ising chain can be
retrieved from the current system. Here we consider the following aspects,
utilizing three quantities, local magnetization, its slope, and local
density of states, to characterize the ground-state properties.

(i) Magnetizations

In a uniform Ising chain, the magnetization of the ground state is uniform.
As the first derivative of the groundstate energy density with respect to
the transverse field, the magnetization experiences a jump at the transition
point in the thermodynamic limit. For a finite chain, it is smooth but
experiences a drastic change near the transition point. According to the
above analysis, this can be observed from the position-dependent
magnetization $\left\langle \sigma _{j}^{z}\right\rangle $\ for the
non-uniform Ising chain with nonzero $g$. Based on the Hellmann-Feynman
theorem, $\left\langle \sigma _{j}^{z}\right\rangle $\ can be computed by 
\begin{equation}
\left\langle \sigma _{j}^{z}\right\rangle =\frac{\partial E_{g}}{%
g_{N}\partial \xi _{j}^{2}},  \label{sigmaz}
\end{equation}%
where $E_{g}$\ is the energy of the ground state and can be obtained from
the exact diagonalization of the corresponding Majorana lattice. The
numerical results of $\left\langle \sigma _{j}^{z}\right\rangle $\ for
different $N$ are shown in Fig.~\ref{fig:fig3}(b). It can be observed that
the bell-shaped curves for different $N$ intersect at the same points $%
\xi_{m_{\pm}}$, and the crossing becomes steeper as $N$ increases. Therefore, in the thermodynamic limit, $\left\langle \sigma _{j}^{z}\right\rangle$ is expected to approach a nearly vertical crossing at $\xi_{m_{\pm}}$.

(ii) The magnetic susceptibility. In a uniform Ising chain, the magnetic
susceptibility\ of the ground state is uniform. As the second derivative of
the groundstate energy density with respect to the transverse field, the
magnetic susceptibility experiences a divergence at the transition point in
the thermodynamic limit. For a finite chain, it is smooth but experiences a
sharp maximum near the transition point. According to the above analysis,
this can be observed from the position-dependent magnetic susceptibility $%
\chi _{j}^{z}$\ for the non-uniform Ising chain with nonzero $g$. Based on
the Hellmann-Feynman theorem, $\chi _{j}^{z}$\ can be computed by 
\begin{equation}
\chi _{j}^{z}=\frac{\partial \left\langle \sigma _{j}^{z}\right\rangle }{%
g_{N}\partial \xi _{j}^{2}},  \label{chiz}
\end{equation}%
which is based on the function of $\left\langle \sigma _{j}^{z}\right\rangle 
$. The numerical results of $\chi _{j}^{z}$\ for different $N$ are shown in
Fig. \ref{fig:fig3}(c). It can be observed that as $N$ increases, the peak
of $\chi_{j}^{z}$ becomes significantly sharper and shifts toward $%
\xi_{m_{\pm}}$. Therefore, in the thermodynamic limit, $\chi_{j}^{z}$ is
expected to diverge at $\xi_{m_{\pm}}$, signaling a phase transition.

(iii) Local density of states. For the uniform system $(g=0)$ studied here,
the ground state is governed by the magnitude of $\delta $. In the
thermodynamic limit, the two phases are separated by a gapless point. This
can be observed from the density of states of the corresponding spectrum of
the Majorana lattice for different values of $\delta $. There exists a
window with vanishing density of states when $\delta \neq \pm 1$. According
to the above analysis, this can be observed from the local density of states
for the Ising chain with nonzero $g$, which is defined as

\begin{equation}
D_{j}=\lambda \frac{\mathrm{d}\mathcal{N}_{j}}{\mathrm{d}\varepsilon },
\label{D_j}
\end{equation}%
where $\mathrm{d}\mathcal{N}_{j}=\sum_{\psi ,\sigma =A,B}\left\vert \langle
\psi |j\rangle _{\sigma }\right\vert ^{2}$, $|\psi \rangle $ denotes the
eigenstates with energies in the interval $[\varepsilon ,\varepsilon +%
\mathrm{d}\varepsilon ]$, and the normalization factors $\lambda =\Delta
\varepsilon /(2N)$ and $\Delta \varepsilon $ is the bandwidth of the
Majorana lattice. Here, $j$ is the unit cell index $j=-N,-N+1,\cdots ,N.$ We
compute $D_{j}$ by using the method \cite{Haydock1980} and the numerical
results for different $N$ are shown in Fig.~\ref{fig:fig3}(d-f). We can see
that the plots of $D_{j}$ become more continuous as $N$ increases. The band
gap reaches its minimum at points $\protect\xi _{m_{\pm }}$, and meanwhile, the mid-gap
modes appear. 

\begin{figure*}[tbp]
	\centering
	\includegraphics[width=0.85\linewidth]{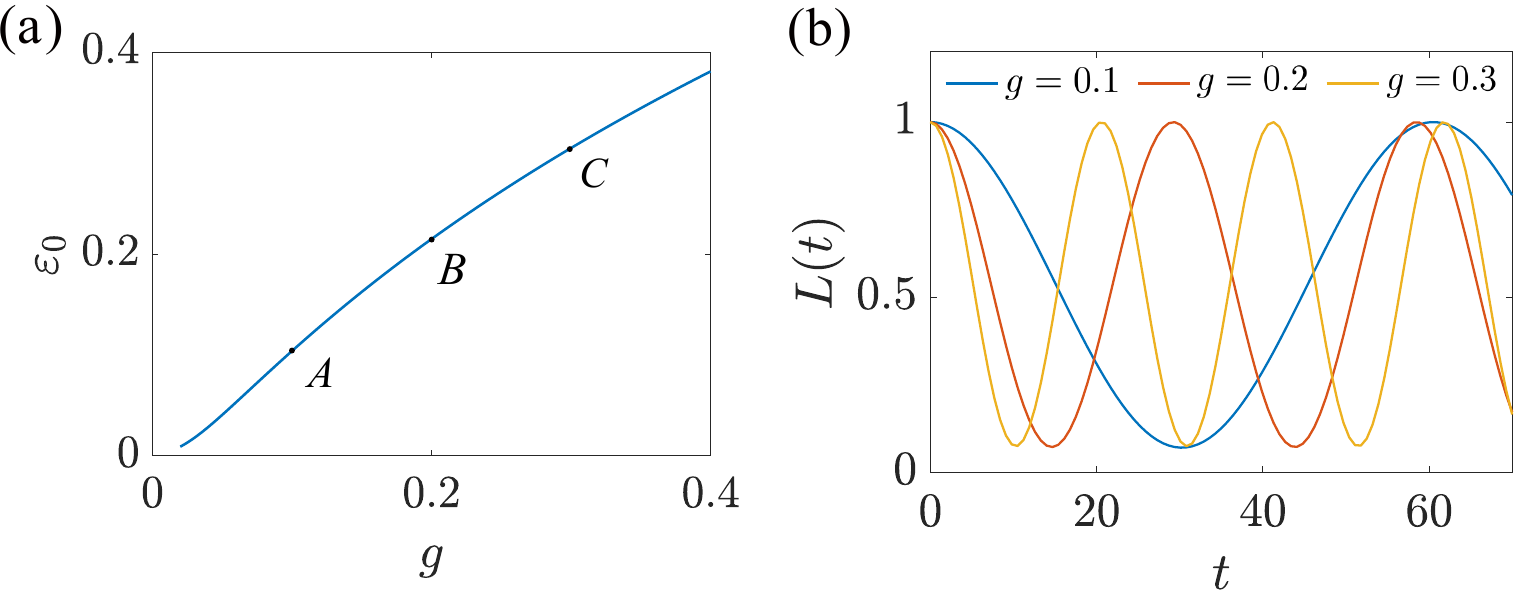}
	\caption{Plots of the energy splitting $\protect\varepsilon _{0}$ in
		Hamiltonian~(\protect\ref{H}) and the Uhlmann-Jozsa fidelity $L(t)$ defined
		in Eq.~(\protect\ref{L_t}). (a) It can be observed that the energy levels
		are quasi-degenerate when the quadratic transverse field strength $g$ is
		small, and the energy splitting increases as $g$ grows. (b) The quench
		dynamics under parameters $A(g=0.1)$, $B(g=0.2)$, and $C(g=0.3)$ in (a). The
		initial state $\protect\rho_0$ is the thermal state of $H_\mathrm{Pre}$ at
		temperature $1/\protect\beta$, as defined in Eq.~(\protect\ref{rho_0}). It
		can be seen that the oscillation periods are respectively $2\protect\pi/%
		\protect\varepsilon _{0}$, which can be exploited to detect the small energy
		splitting. Other parameters: $N=5$, $\protect\delta =0.5$, and $J=1$.}
	\label{fig:fig4}
\end{figure*}

\section{Thermal state oscillation}

\label{Thermal state oscillation}

The above analysis shows that the quadratic transverse-field Ising spin
chain has peculiar properties. Besides the expected profiles of several
observables along the chain, the fascinating feature is the quasi-degeneracy
of the spectrum. In this section, we focus on the dynamical demonstration of
such a quasi-degenerate spectrum. Our strategy is to employ quench dynamics
starting from a thermal state to detect the small energy splitting $%
\varepsilon _{0}$. To proceed, the pre-quench Hamiltonian and the
post-quench Hamiltonian\ are defined as%
\begin{eqnarray}
H_{\mathrm{Pre}} &=&H+\kappa (d_{0}+d_{0}^{\dag }),  \notag \\
H_{\mathrm{Pos}} &=&H.  \label{HpreHpos}
\end{eqnarray}%
Based on Eq.~(\ref{DPsi}), in the full basis set of eigenstates $\left\{
|\psi _{n}^{+}\rangle ,|\psi _{n}^{-}\rangle \right\} $, the Hamiltonian $H_{%
\mathrm{Pre}}$\ can be expressed in the block-diagonal form of a $2\times 2$
matrix%
\begin{equation}
h_{n}=\frac{E_{n}^{-}+E_{n}^{+}}{2}+\left( 
\begin{array}{cc}
\varepsilon _{0}/2 & \kappa \\ 
\kappa & -\varepsilon _{0}/2%
\end{array}%
\right) .
\end{equation}%
The eigenstates of $h_{n}$ are 
\begin{eqnarray}
|\varphi _{n}^{+}\rangle &=&\sin \theta \,|\psi _{n}^{+}\rangle +\cos \theta
\,|\psi _{n}^{-}\rangle  \notag \\
|\varphi _{n}^{-}\rangle &=&\cos \theta \,|\psi _{n}^{+}\rangle -\sin \theta
\,|\psi _{n}^{-}\rangle
\end{eqnarray}%
with eigenenergy $\mathcal{E}_{n}^{\pm }=\left( E_{n}^{-}+E_{n}^{+}\right)
/2\pm \sqrt{\varepsilon _{0}^{2}/4+\kappa ^{2}}$, where $\tan \left( 2\theta
\right) =2\kappa /\varepsilon _{0}$. In the case with $\kappa \gg
\varepsilon _{0}$, we have

\begin{equation}
|\varphi _{n}^{\pm }\rangle \approx \frac{|\psi _{n}^{+}\rangle \pm |\psi
_{n}^{-}\rangle }{\sqrt{2}}.
\end{equation}%
The initial state is taken as the thermal state at temperature $1/\beta $ of 
$H_{\mathrm{Pre}}$, given by the density matrix%
\begin{eqnarray}
\rho (0) &=&\frac{1}{Z}e^{-\beta H_{\mathrm{Pre}}}  \notag \\
&=&\frac{1}{Z}\sum_{n,\sigma =\pm }\exp (-\beta \mathcal{E}_{n}^{\sigma
})|\varphi _{n}^{\sigma }\rangle \langle \varphi _{n}^{\sigma }|,
\label{rho_0}
\end{eqnarray}%
with $Z=\mathrm{Tr}\left( e^{-\beta H_{\mathrm{Pre}}}\right) $. To capture
the effect of $\varepsilon _{0}$, we compute the time evolution of the state
by

\begin{eqnarray}
\rho (t) &=&\sum_{n,\sigma =\pm }\exp (-\beta \mathcal{E}_{n}^{\sigma
})e^{-iHt}|\varphi _{n}^{\sigma }\rangle \langle \varphi _{n}^{\sigma
}|e^{iHt}  \notag \\
&=&\sum_{n,\sigma =\pm }\exp (-\beta \mathcal{E}_{n}^{\sigma })|\varphi
_{n}^{\sigma }(t)\rangle \langle \varphi _{n}^{\sigma }(t)|.
\end{eqnarray}%
Here $|\varphi _{n}^{\sigma }(t)\rangle =\left( |\psi _{n}^{+}\rangle
+\sigma \exp (i\varepsilon _{0}t)|\psi _{n}^{-}\rangle \right) /\sqrt{2}$ is
periodic with period $2\pi /\varepsilon _{0}$ for every different $n$ and $%
\sigma $. Accordingly, $\rho (t)$ is also periodic with period $2\pi
/\varepsilon _{0}$. To measure this feature, we employ the Uhlmann-Jozsa
fidelity, given by 
\begin{equation}
L(t)=\left[ \mathrm{Tr}\sqrt{\sqrt{\rho (0)}\rho (t)\sqrt{\rho (0)}}\right]
^{2}.  \label{L_t}
\end{equation}%
To verify and demonstrate the above analysis, numerical simulations are
performed for finite-size lattices with several typical parameters by exact
diagonalization. We plot the energy splitting $\varepsilon _{0}$ and the
fidelity $L(t)$\ in Fig. \ref{fig:fig4}. These numerical results are
consistent with our analytical predictions. It can be observed that, for a
given $\delta$, the energy levels are quasi-degenerate when the quadratic
transverse field strength $g$ is small, and the energy splitting increases
as $g$ grows. Meanwhile, the oscillation periods $T$ of the thermal state of 
$H_{\mathrm{Pre}}$ at any temperature can be used to obtain the small energy
splitting $2\pi/T$.

\section{Summary}

\label{Summary}

In summary, we have extended the concept of topological degeneracy emerging
in a uniform open chain to the one with a quadratic transverse field. In
comparison with the uniform chain, where topological degeneracy only exists
in the nontrivial phase, the present system always supports the Kramers-like
degeneracy. The underlying mechanism is the existence of localized Majorana
modes residing at the interface of two domains with different quantum
phases. This also results in special distributions of magnetization and
local density of state along the chain. We also show that the Kramers-like
degeneracy is lifted by a constant shift due to the hybridization of the two
localized modes when the trapping field strength becomes stronger. As a
dynamical demonstration, this Zeeman effect leads to periodic oscillations
for a finite-temperature thermal initial state. These findings lay the
foundation for exploring novel physical phenomena induced by trapping fields
in interacting systems.

\acknowledgments This work was supported by NSFC (Grant No.12374461).

\end{document}